# Interferometer observations of pulse pairs in an interstellar communication experiment

*William J. Crilly Jr.*

Green Bank Observatory, West Virginia, USA

*Abstract*—Synchronized radio telescope-based experiments conducted since 2017, together with subsequent interferometer experiments, provide evidence of an anomalous source of 3.7 Hz bandwidth pulses, sourced from near the direction of the star Rigel. The current experiment, reported here, uses a two-element phase-measuring interferometer to monitor the hypothetical pulse source across azimuths within the beam-widths of the elements of a south-facing interferometer. 123 days of phase measurements of 3.7 Hz bandwidth pulse pairs, adds to the prior evidence that the pulsing signal source has celestial origin. Associated measurements of noise power in 954 Hz and 50 MHz bandwidths, made simultaneous with the 0.27 second duration pulse pair measurements, are presented. Measurement results are presented to aid in the development of independent experimental replication, and alternate and auxiliary explanatory hypotheses.

## I. INTRODUCTION

**Field of study**

Humans have speculated that electromagnetic signals may be used to convey information among intelligent civilizations across interstellar distances. The reception of these speculated signals is a difficult undertaking and is confounded by many obstacles. The work presented here is an attempt to follow an evidence path, finding measurement anomalies that may support a hypothesis that such communication signals are present. Extensive future work is required before an interstellar communication hypothesis may be accepted as the most likely explanation of measurement anomalies.

**Prior work**

The initial identification of a near-Rigel hypothetical narrowband transmitting celestial direction resulted from the measurement evidence of simultaneous high signal to noise ratio (SNR) pulses observed using the Forty-Foot Radio Telescope at the Green Bank Observatory in West Virginia, synchronized with the sixty-foot diameter Plishner Radio Telescope of the Deep Space Exploration Society, near Haswell, Colorado.[1] Subsequent work and replication provided evidence supporting a celestial source hypothesis. [2]-[9] There is no independent experimental replication of this work known to the author.

**Objectives**

The experiment reported here examines signal phase measurements to deduce 3.7 Hz bandwidth pulse pair direction of arrival, as the hypothetical transmitting source is located at azimuth angles covering the range of the antenna beam-widths. Such evidence might increase the signal candidate pulse pair flux count, compared to previous work [1]-[9], resulting in increased statistical power supporting a celestial signal source hypothesis. The experiment entails examining whether interferometer phase-based direction of arrival measurements are consistent with the expected changing azimuth direction of a celestial source, during a long duration. If such consistency appears, then a celestial source hypothesis is supported, notwithstanding the existence of alternate and auxiliary hypotheses.

An additional objective is to report measurements that may provide evidence of a natural object explanation. Associated measurements, e.g. pulse pair simultaneous noise power in 954 Hz and 50 MHz bandwidths, per antenna element, provide evidence that may be used to study natural object source hypotheses, e.g. spectral outliers, masers, pulsars, and fast radio bursts. Associated measurements may also help solve some of the difficult problems of identifying Radio Frequency Interference (RFI).

## II. HYPOTHESIS

**Hypothesis**

Conjectured interstellar communication signals may comprise narrow bandwidth, short duration pulses having readily discoverable properties, including repetitive transmissions that permit celestial origin discovery, source identification, high information capacity communications, and distinction compared to RFI and natural objects.

Falsification of the hypothesis is possible in several ways —for example, if a likely RFI explanation emerges, if a natural object model accounts for the anomalies, if equipment is found to be contributing to the anomalies, or if repeated attempts fail to replicate them.

## III. METHOD OF MEASUREMENT

**Summary**

**Figure 1** describes the system used in this work. In summary, two offset-fed paraboloidal reflectors are pointed to scan the -4.3° declination directions, at a fixed azimuth of 180.0°. Four phased folded dipoles, tuned to cover 1400-1450 MHz, are used at each feed. Right hand circular antenna-incident polarized signals are produced, amplified, down-converted and digitized. Measurements of pulses exceeding 8.5 dB in a 3.7 Hz bandwidth are made and recorded in four hour duration files over 123 days. Second level processing is performed on these files to seek anomalous pulse pairs. Details are described in prior reports. **[1]-[9]**

**Observation run O10 differences from O9**

The first-level processed files used in O10, reported here, are the same as those used in O9 **[9]**, while differing in second-level processing.

Salient differences between O9, described in **[9]**, and O10 are:

**1.** The interferometer element phase difference filter, setting limits of **Δ$_{EW}$ϕ,** covers the range of a phase-

# Interferometer observations of pulse pairs in an interstellar communication experiment

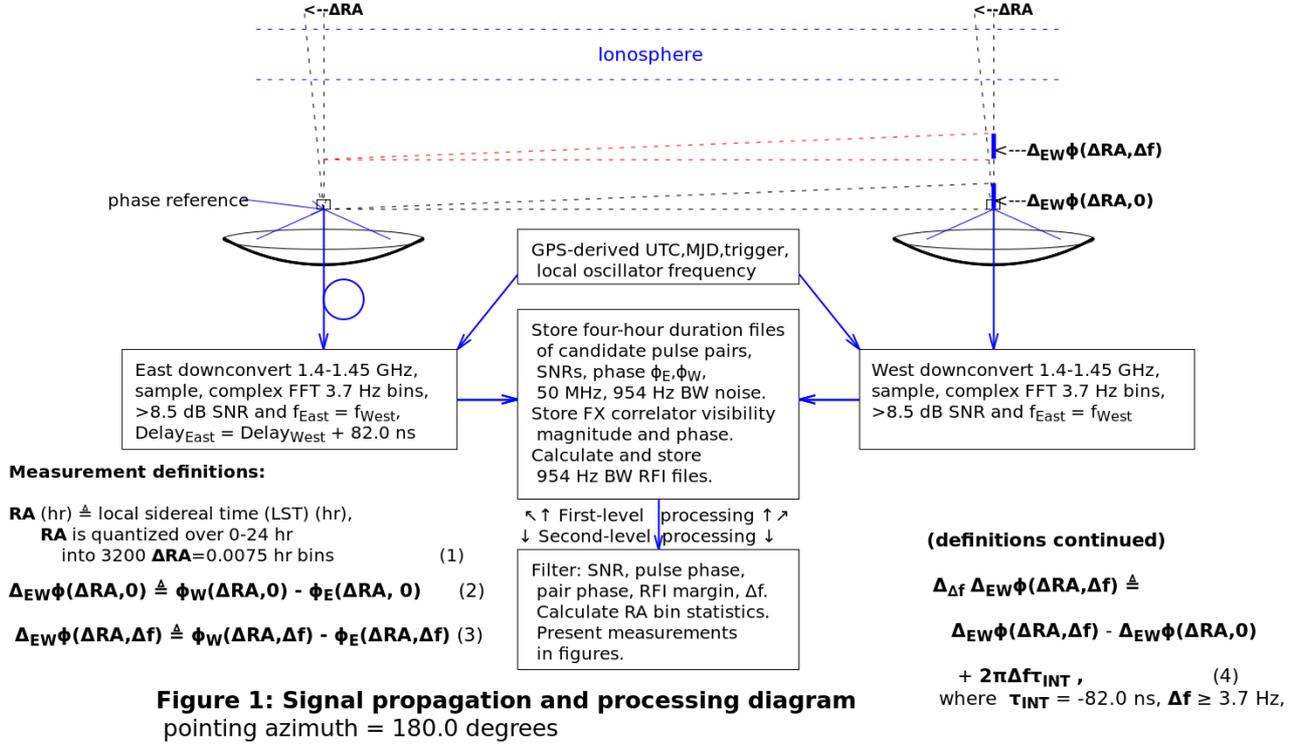

**Figure 1:** The interferometer-based phase measurement system processes In-phase and Quadrature (IQ) baseband signals to provide evidence indicating a common direction of arrival of two simultaneous 3.7 Hz bandwidth pulses, separated in the range of $\Delta f$ = 300 to 540 kHz, in the experiment reported here. System details are described in previous reports. [1]-[9]

wrapped measurement, i.e. $\pm\pi$. In prior work the filter was set to a low limit range to seek pulse pairs propagating from a common narrow azimuth direction. In this work, the limit is opened in order to capture signals arriving at angles within the beam-width of the interferometer elements. The full range phase measurement allows one to examine $\Delta_{EW}\phi$ measurements relative to telescope measured RA pointing at the time of the pulse pair acquisition, deducing a common celestial direction of arrival during the experiment, over a wider azimuth range.

**2.** The $\Delta_{EW}\phi$ measurement is phase wrapped to $\pm\pi$ from $\pm 2\pi$. The $\pm 2\pi$ phase distribution is triangular, resulting from the convolution of the $\pm\pi$ uniform distributions of the measured element phase values. The triangular distribution complicates the binomial pulse pair count calculation due to variable event probabilities. After phase wrapping, pulse pairs have an event probability in Additive White Gaussian Noise (AWGN) independent of $\Delta_{EW}\phi$. Some ambiguity may exist near the $\pm\pi$ and $\pm 2\pi$ boundaries, as phase noise might change the sign of the wrapped $\Delta_{EW}\phi$ measurement. Phase noise causes are described in **Theory of Operation** in **[8]**.

**3.** An anomalous concentration of pulse pairs measuring $\Delta f$ approximately between 300 and 540 kHz was observed in the near-Rigel Direction Of Interest (DOI). Pulse pairs are filtered to be in this $\Delta f$ range.

**4.** The pulse pair $\Delta_{\Delta f}\Delta_{EW}\phi$ filter was reduced from $\pm 0.8$ radians to $\pm 0.18$ radians to seek pulses closer to the expected center of the conjectured phase distribution.

**5.** SNR is filtered using the composite of four pulse SNR, antenna element {East,West} and pulse pair {0,$\Delta f$}.

**Ionospheric effects**

Given that the pulse pair RF frequencies in this work are much greater than the atmospheric plasma frequency and cyclotron frequency, the atmosphere is effectively transparent, albeit having scattering effects due to rain and snow. Other remaining effects include delay and polarization changes due to ionospheric charge and the Earth's magnetic field. **[10]**

While the magnetic field is perpendicular to the direction of propagation, a right hand circular polarized wave will have a calculated phase shift due to Faraday Rotation of

$$\phi = 2.36 \times 10^{-14} \ B \ N_t / f_0^2 \text{ radians,} \quad (5)$$

where $B$ is the magnetic field strength in Teslas, $N_t$ is the Total Electron Count (TEC) in electrons per square meter, and $f_0$ is RF frequency in GHz. **[11][12]**

A prediction of the Faraday Rotation induced phase shift difference between two simultaneous pulses, with the second pulse offset from frequency $f_0$ by $\Delta f$, may be calculated from a first derivative of equation **(5)**, with respect to $f_0$, giving

$$\Delta \phi = -4.72 \times 10^{-17} \, B \, N_t \, \Delta f / f_0^3 \text{ radians}, \quad (6)$$

where $\Delta f$ is in MHz.

An example calculation at an Earth average magnetic field strength $B$ of 50 microTeslas, $f = 1.425$ GHz and a relatively high value of electron density at $N_t = 10^{18}$ electrons per m$^2$ **[11]**, estimates a phase shift between simultaneous pulses in the pair at $8.1 \times 10^{-4}$ radians per MHz of frequency difference. The maximum $\Delta f$ used in the O10 filtered data set is 0.54 MHz, while the $\Delta_{\Delta f}\Delta_{EW}\phi$ filter is set to $0 \pm 0.18$ radian. It is therefore expected that the difference in the phase between the two pulses in each candidate pulse pair due to Faraday Rotation is significantly less than the $\Delta_{\Delta f}\Delta_{EW}\phi$ filter range, and is not expected to affect the statistical power of the measurement.

The rate of change of TEC has been observed in the auroral zone, using satellite measurements, at a maximum of $0.7 \times 10^{16}$ electrons per square meter per second. **[11]** Using equation **(5)**, during an FFT integration interval, a phase shift of $1.6 \times 10^{-3}$ radians is expected, and may be neglected.

Propagation time delay may be estimated using

$$t = 1.345 \times 10^{-19} \, N_t / f_0^2 \text{ microseconds}, \quad (7)$$

where $f_0$ is frequency in GHz, and $N_t$ is the electron density along the propagation path.**[11]** The delay at frequency $f_0 = 1.425$ GHz and $N_t = 10^{18}$ electrons/m$^2$ calculates to 66 nanoseconds. The delay difference across interferometer elements depends on the variation in electron density across the first Fresnel zones in a Fresnel screen model. This variation may be estimated using an ionospheric refraction model. A maximum likely refraction value is expected to be 0.05º at 100 MHz, from **[12]**, Table 13.6. Using this value, a negligible expected phase difference across 33 wavelength spaced interferometer elements, is calculated at 1425 MHz to be $8.9 \times 10^{-4}$ radians.

In summary, the high transparency of the ionosphere at 1398 to 1451 MHz is expected to permit accurate measurement of narrowband signal phase differences, providing angle of arrival measurement of simultaneous constituent pulses in pulse pairs, during a long duration experiment.

**RFI Considerations**

Suspected RFI identification and excision are implemented in machine algorithms primarily based on the population of narrow band pulses that appear in spectral segments of 256 FFT bins, spanning 954 Hz. Details of RFI amelioration methods are contained in **[1]-[5]** and **[7]-[9].**

**Multiple experimental methods**

An important aspect of experiments is the replication of prior observations using differing experimental methods. Prior work has utilized three experimental methods:

1. Synchronized widely-spaced radio telescopes, **[1]**
2. Beam transits of a standalone radio telescope, **[2]-[4]**
3. Interferometer measurements of a common direction of arrival of $\Delta f$ separated pulses in pulse pairs, using a south-facing interferometer with phase measurement filters sensitive to a narrow arrival angle. **[5]-[9]**

In the current work, the experiment deduces the angle of arrival of $\Delta f$-separated simultaneous pulses in pairs, across the antenna element beam-width, while comparing the deduced angle of arrival to that predicted from a celestial source.

## IV. OBSERVATIONS

A 123 day data set was processed, in second level processing, to seek anomalies in near-Rigel pulse pair measurements. DOIs identified in prior work, other than the near-Rigel direction, are not reported in figures here.

Observations are arranged in three groups.

**Figs. 2-4** plot measurements of near-Rigel pulse pair anomalies showing phase measurement and statistical power of pulse pair counts.

**Figs. 5-8** plot measurements that show wider bandwidth power spectral density simultaneous with the acquisition of pulse pairs, so that the presence of natural objects may be gleaned.

**Figs. 9-11** plot associated measurements that aid in understanding the anomalies and in seeking RFI and/or natural object explanations.

Details follow:

**Fig. 2** plots the interferometer phase of the first pulse in a pulse pair, against the MJD-dependent pointing RA of 180.0º azimuth perpendicular to baseline pointing. A line is included in the figure that plots the expected phase measurement of an object having a celestial coordinate RA at the central DOI value, given a 0.116 hr fringe period due to the 33.0 wavelength baseline distance, at 1425 MHz. Lines parallel to this line are plotted at spacings of approximately two RA bins. This figure is useful to glean evidence of pulse pair transmissions originating from a celestial source.

**Figs. 3-4** plot the effect size of pulse pair count across RA bins and at different levels of SNR threshold. This figure is useful to determine the statistical power of the repetition of prior observations.

**Figs. 5-8** plot noise measurements in 954 Hz and 50 MHz bandwidth for each interferometer element, made at the time of each pulse pair acquisition. This aids the development of natural object model hypotheses. For example, high continuum emissions from natural objects has a potential to produce anomalous narrow bandwidth pulse pairs, given Rayleigh statistics, and leakage of false positives due to power fluctuations and non-AWGN signal properties. The absence of wide bandwidth spectral power density measurement reduces the likelihood of such false positive leakage effects.

**Figs. 9-10** plot the MJD and RF Frequency in MHz of the acquired filtered pulse pairs. RFI sometimes appears repetitively, concentrated in time and RF frequency. RFI protected spectrum may be examined. A high capacity interstellar communication transmitter is expected to have different characteristics compared to much RFI, given a conjecture that simultaneous transmissions on a wide range of frequencies is implemented in the transmitter.

**Fig. 11** plots the frequency spacing $\Delta f$ of the simultaneous pulses in each pulse pair. Concentrations of pulse pairs in a range of values is unexpected in a natural object having AWGN-like power spectral density. RFI is not expected to have such a concentration, together with a common celestial direction, during a 123 day experiment.

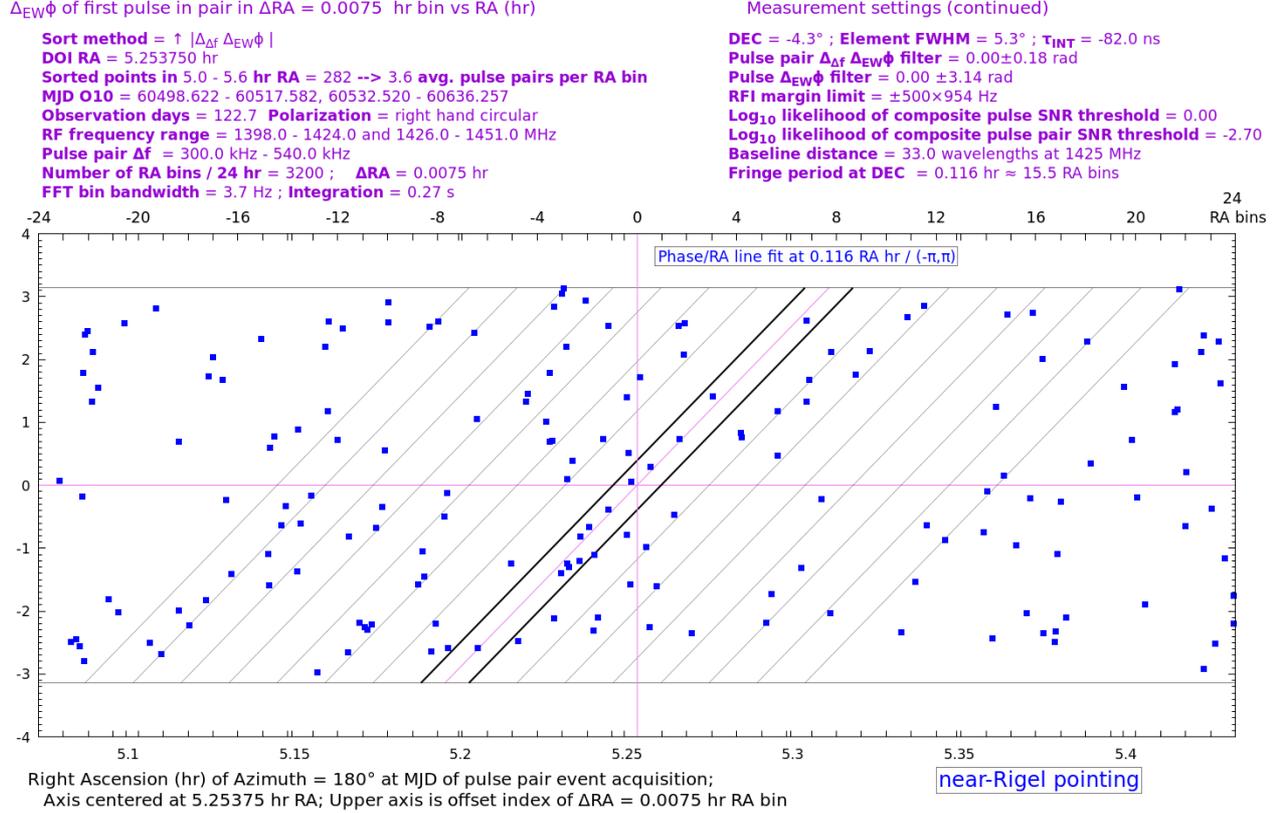

**Figure 2:** The phase difference between the east and west interferometer elements of the first pulse in a pulse pair is plotted against the RA of the interferometer baseline perpendicular pointing at 180º azimuth, i.e. the telescope Local Sidereal Time (LST). This LST is derived from the GPS-time derived MJD recorded by one of the receiver system signal digitizers at the time of the pulse pair acquisition. The expected phase/RA slope is expected to be $2\pi$ radians per 0.116 hr RA, explained as follows.

Pulse pairs from a hypothetical celestial source, directionally located east of south, i.e. having a celestial RA greater than the GPS/MJD-derived RA, i.e. the LST, will be recorded with a lower RA than that of the hypothetical easterly object, and will measure an increasing phase shift across antenna elements, over time, resulting from the propagation angular direction offset. The slope of the line is $2\pi$ radians per 0.116 hr RA, the latter value being the interferometer fringe period, calculated from the measured element spacing along the baseline, declination pointing, and the wavelength at 1425 MHz. The central RA value, 5.25375 hr RA, was found after searching the 5.1 to 5.4 hr RA range for high statistical power of LST-concentrated pulse pair counts. Phase measurements aid in identifying RFI.

The average number of pulse pairs per RA bin in the 5.1-5.6 hr RA range is measured at 3.6. Thirteen pulse pairs were observed in the diagonal region of the near-Rigel DOI. RA bin traversal of a celestial object calculates to 27 seconds in time.

RFI would generally not be expected to be synchronized with this narrow (27 s) celestial time window during 123 days. Further, RFI would tend to not have the celestial predicted phase measurement values of each of the four constituent pulses of each pulse pair acquired by the two elements of the interferometer.

Interferometer observations of pulse pairs in an interstellar communication experiment

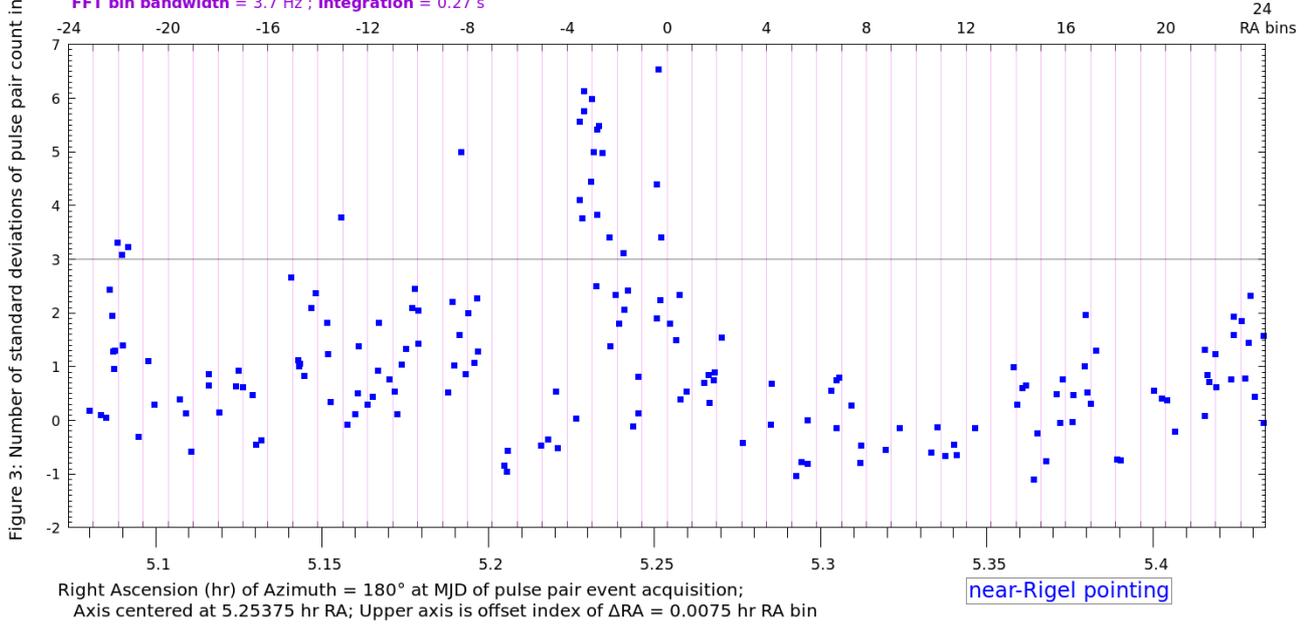

**Figure 3 : Observation run O10 Δt=0 Δf polarized pulse pair measurement:**
Number of standard deviations of pulse pair in ΔRA = 0.0075 hr bin vs RA (hr)

Measurement settings (continued)

**Sort method** = ↑ |Δ_Δf Δ_EW φ |
**DOI RA** = 5.253750 hr
**Sorted points in 5.0 - 5.6 hr RA** = 282 --> 3.6 avg. pulse pairs per RA bin
**MJD O10** = 60498.622 - 60517.582, 60532.520 - 60636.257
**Observation days** = 122.7  **Polarization** = right hand circular
**RF frequency range** = 1398.0 - 1424.0 and 1426.0 - 1451.0 MHz
**Pulse pair Δf** = 300.0 kHz - 540.0 kHz
**Number of RA bins / 24 hr** = 3200 ;   **ΔRA** = 0.0075 hr
**FFT bin bandwidth** = 3.7 Hz ; **Integration** = 0.27 s

**DEC** = -4.3° ; **Element FWHM** = 5.3° ; **τ_INT** = -82.0 ns
**Pulse pair Δ_Δf Δ_EW φ filter** = 0.00±0.18 rad
**Pulse Δ_EW φ filter** = 0.00 ±3.14 rad
**RFI margin limit** = ±500×954 Hz
**Log_10 likelihood of composite pulse SNR threshold** = 0.00
**Log_10 likelihood of composite pulse pair SNR threshold** = -2.70
**Baseline distance** = 33.0 wavelengths at 1425 MHz
**Fringe period at DEC** = 0.116 hr ≈ 15.5 RA bins

near-Rigel pointing

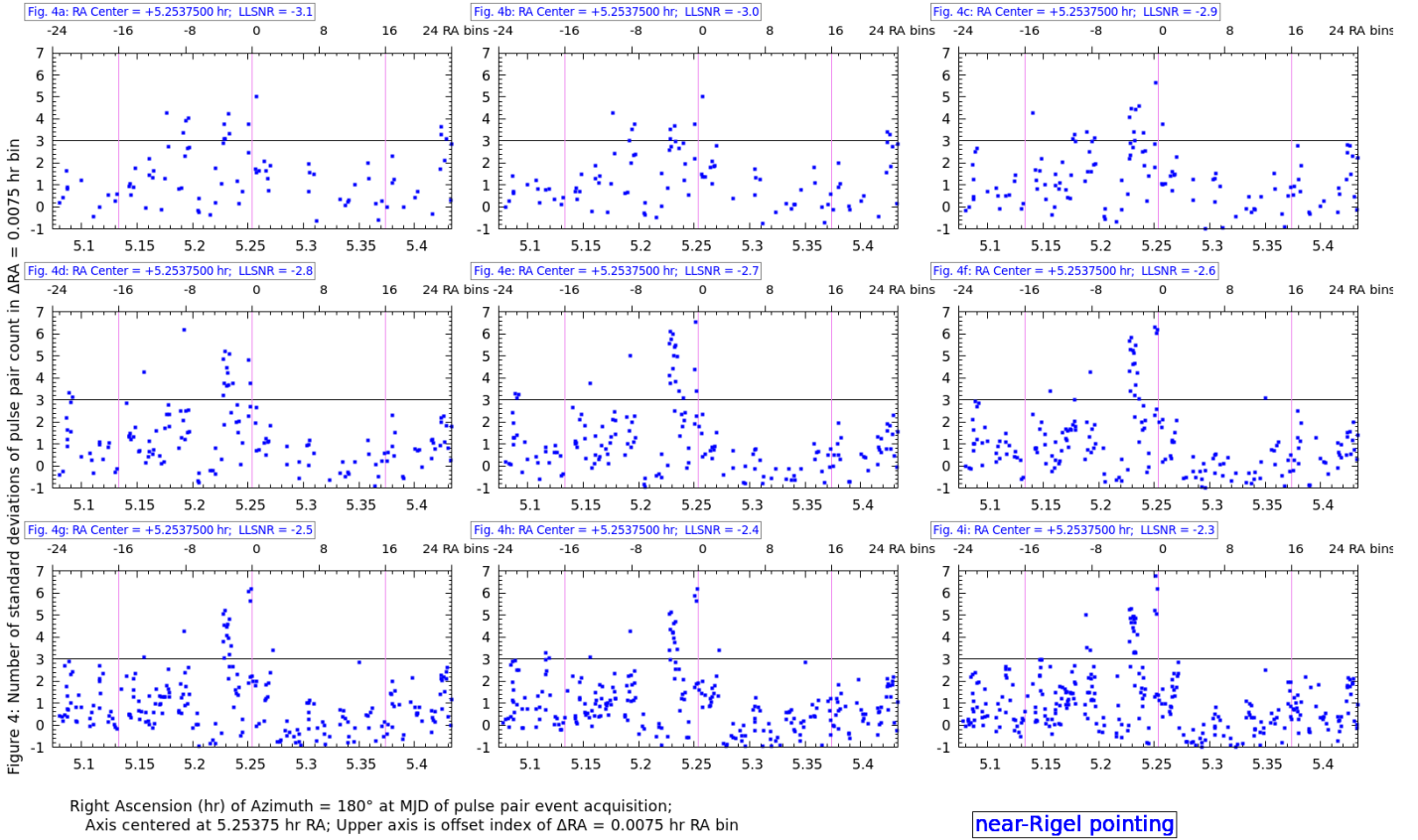

near-Rigel pointing

Right Ascension (hr) of Azimuth = 180° at MJD of pulse pair event acquisition;
Axis centered at 5.25375 hr RA; Upper axis is offset index of ΔRA = 0.0075 hr RA bin

**Figures 3-4:** The number of standard deviations of the mean shift of the pulse pair count is plotted, at the $Log_{10}$ Likelihood of SNR (LLSNR) equal to the central value of -2.7, in **Fig. 3.** and in a range of LLSNR settings, in **Fig. 4**. LLSNR is defined in **Equ. (2)** in **[7],** with the sum over elements and pulses in the pair. Mean shift measured in standard deviations helps determine statistical power, measuring Cohen's d effect size across a population. **[13]** High concentrations of pulse pairs are indicated in the direction of the near-Rigel DOI, observed in prior work. **[1]-[9]**

Interferometer observations of pulse pairs in an interstellar communication experiment

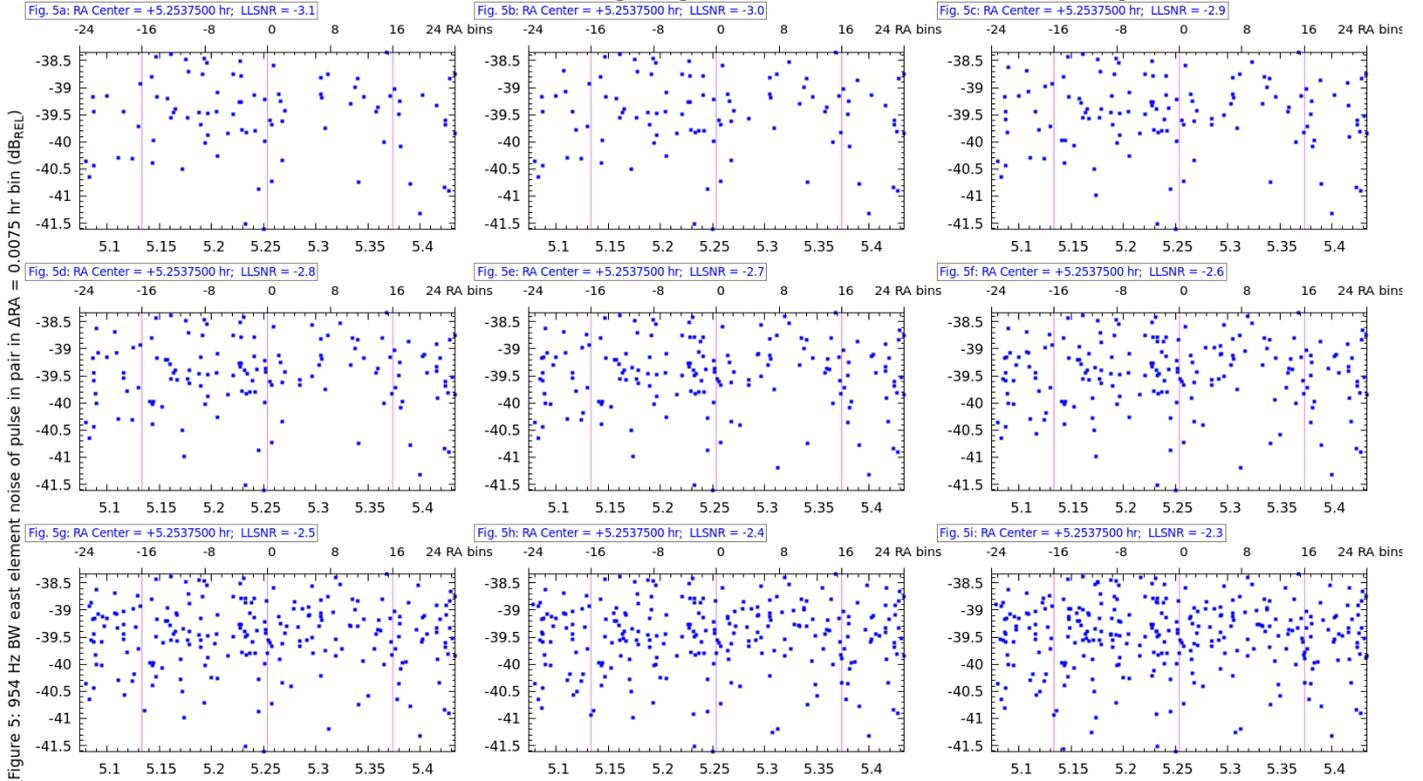

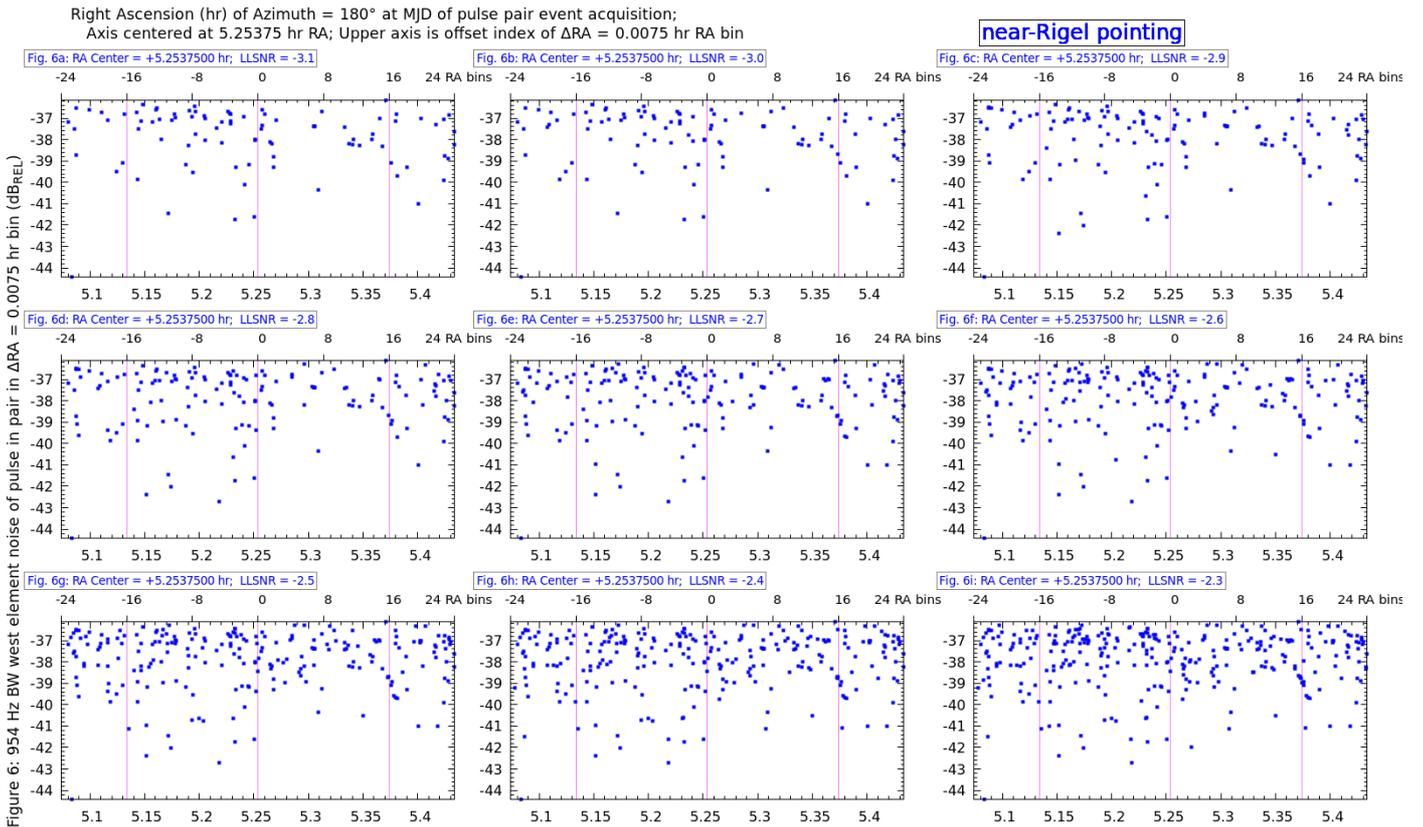

**Figures 5-6:** The east and west element SNR values of pulses are calculated in first-level processing using a 3.7 Hz bandwidth power measurement and a 954 Hz bandwidth noise measurement, the latter normalized to 3.7 Hz bandwidth. The 954 Hz spectral segment contains the 3.7 Hz bandwidth signal. The noise measurement is plotted above at changing LLSNR threshold. Given an anomalous measurement of a >3.7 Hz bandwidth noise burst, one might expect to see an increase in the 954 Hz bandwidth power measurement. The absence of this increase implies that the 3.7 Hz bandwidth high SNR measurement is not the result of a significantly wider bandwidth emission. In addition, emissions that have spectral content comprising a portion of the 954 Hz bandwidth might tend to produce a large number of simultaneous pulse pairs. A search for associated wide band emissions is ongoing work.

Interferometer observations of pulse pairs in an interstellar communication experiment

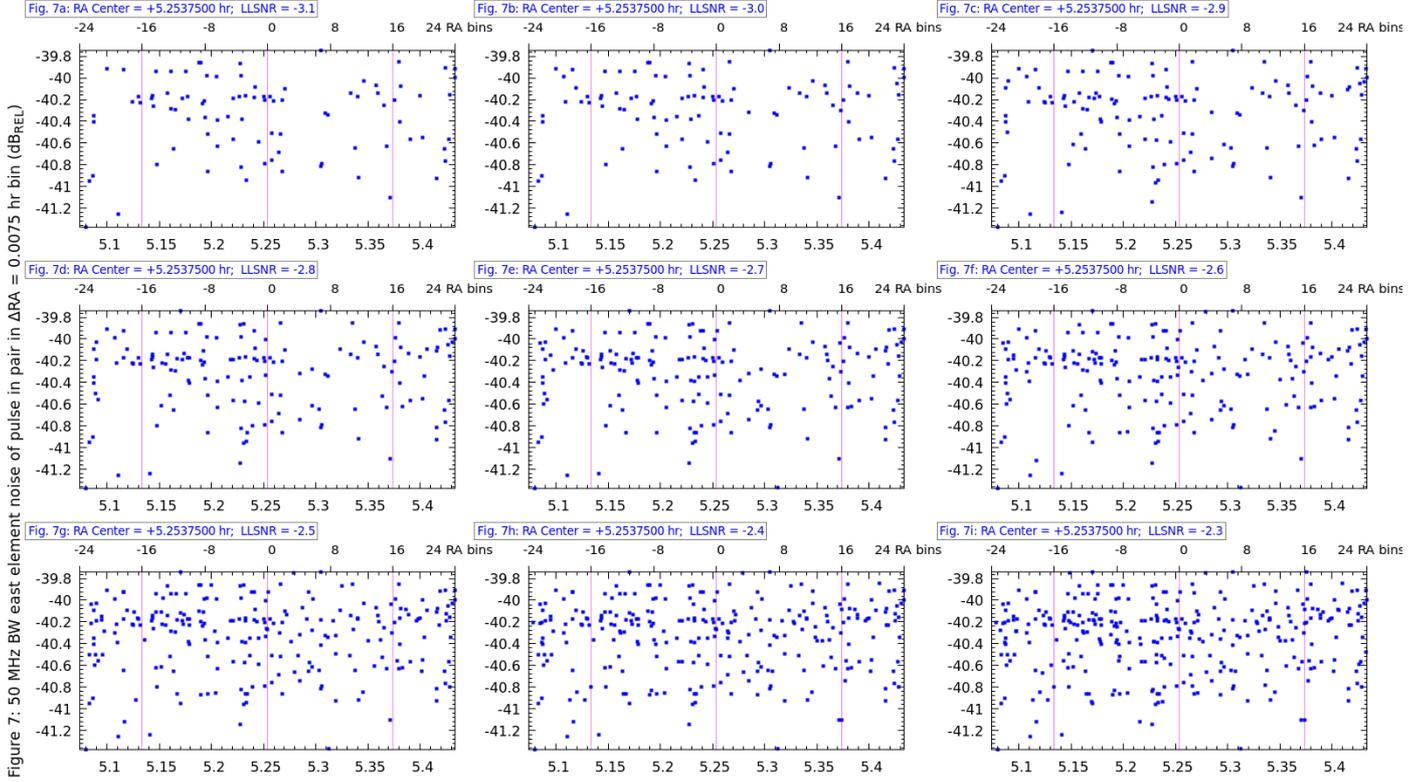

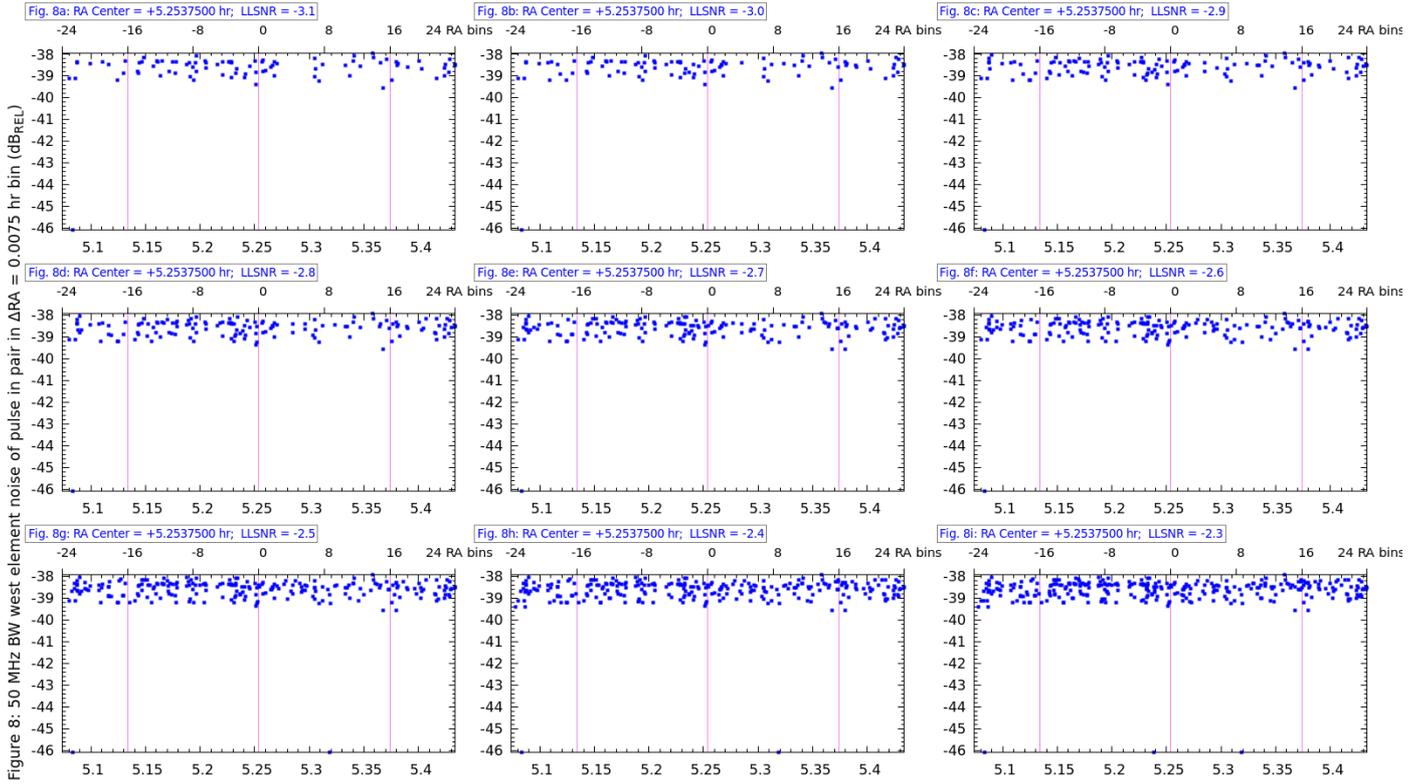

**Figures 7-8:** The 50 MHz noise power measurement allows examination of possible wide bandwidth noisy signals received at the time of the pulse pair acquisition. Such increased measurements do not appear in the direction of interest. This measurement implies that the pulse pairs are not coincident with a wide bandwidth emission. In general, the presence of a narrow bandwidth signal, absent wide band emissions, might be indicative of either a highly compact, high power narrowband emission, or a large homogeneous surface emitter having a low rotation rate, to minimize Doppler spread. Three low levels of west element measurement in **Fig. 8** have unknown cause.

Interferometer observations of pulse pairs in an interstellar communication experiment

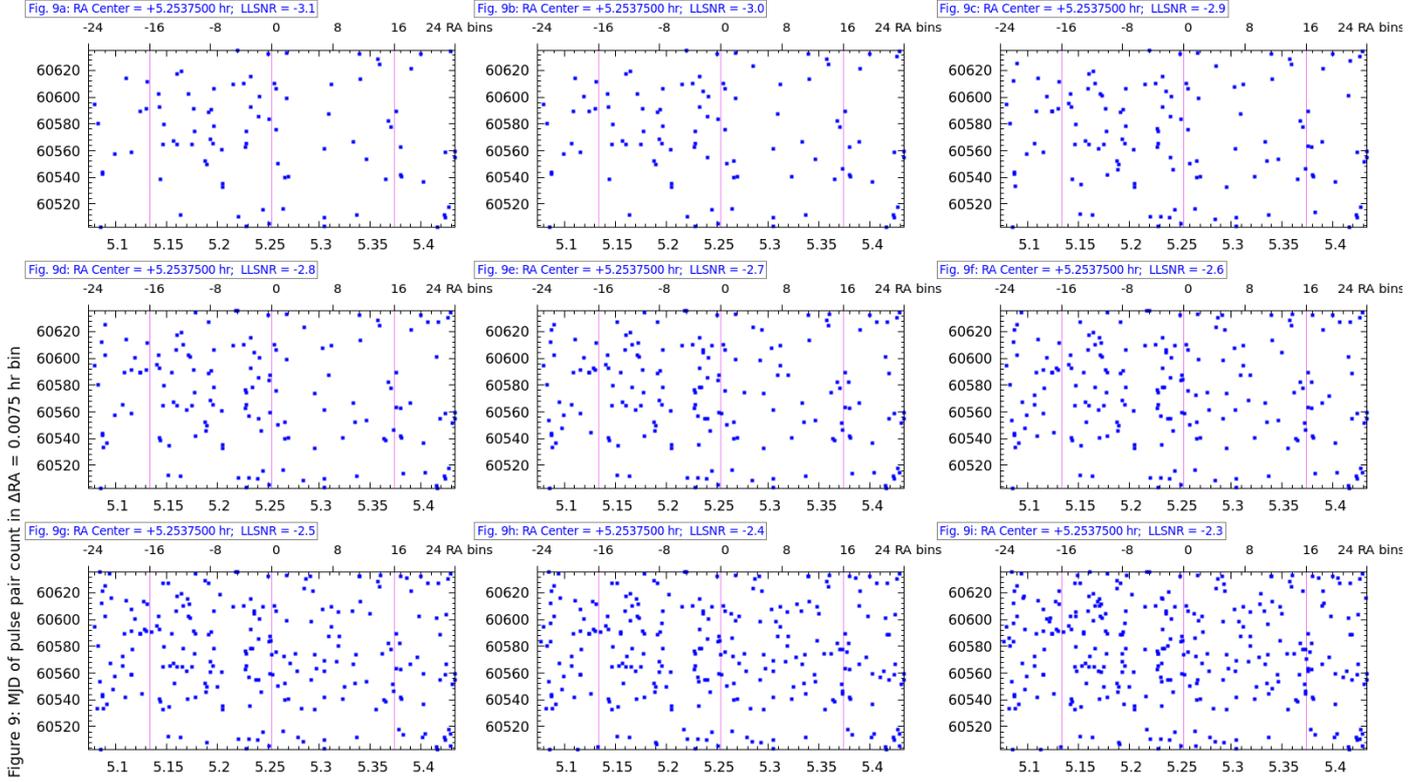

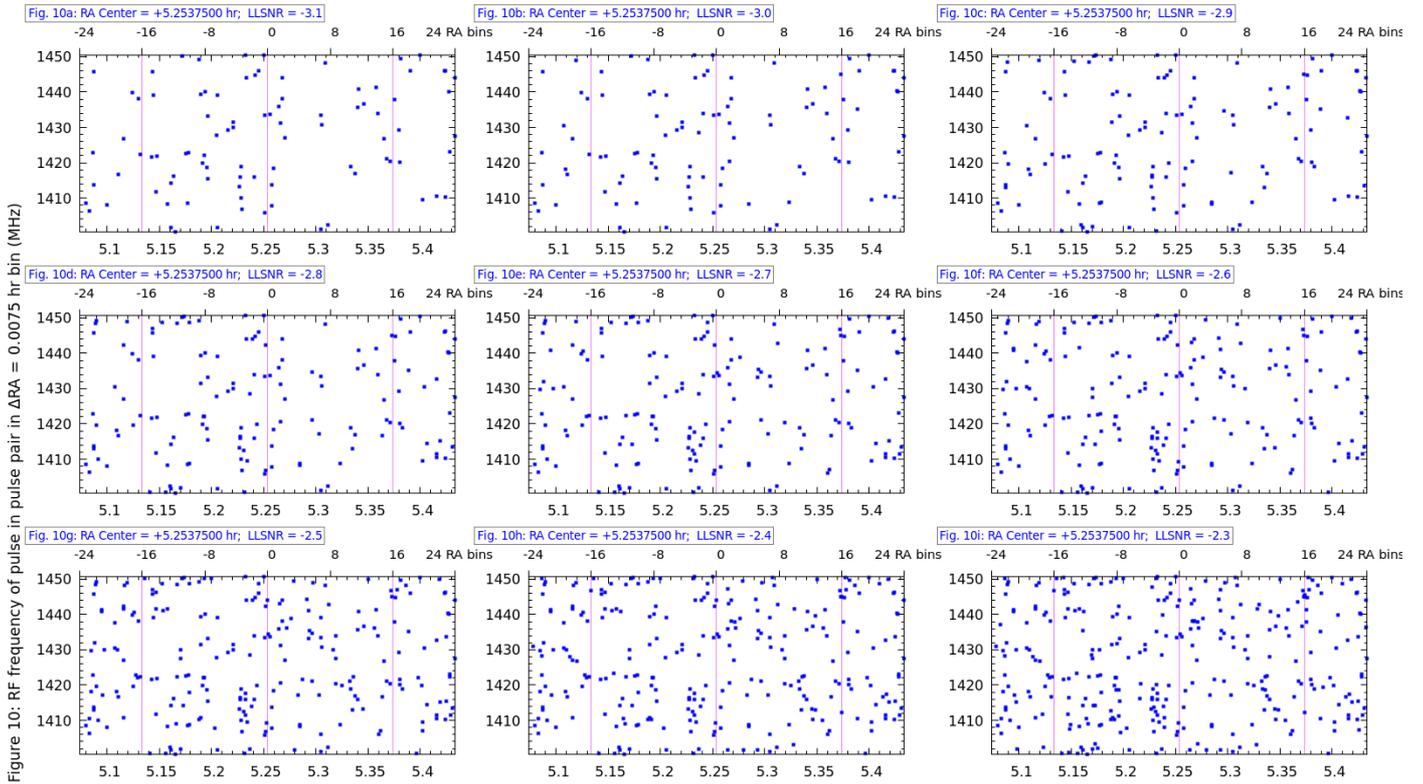

**Figures 9-10:** The MJD and RF Frequency of the first pulse in each pulse pair are plotted at various levels of LLSNR. A concentration in the range below 1425 MHz may be due to a telescope sensitivity issue. The 1400-1427 MHz range is an internationally protected band, limiting emissions. **[14]**

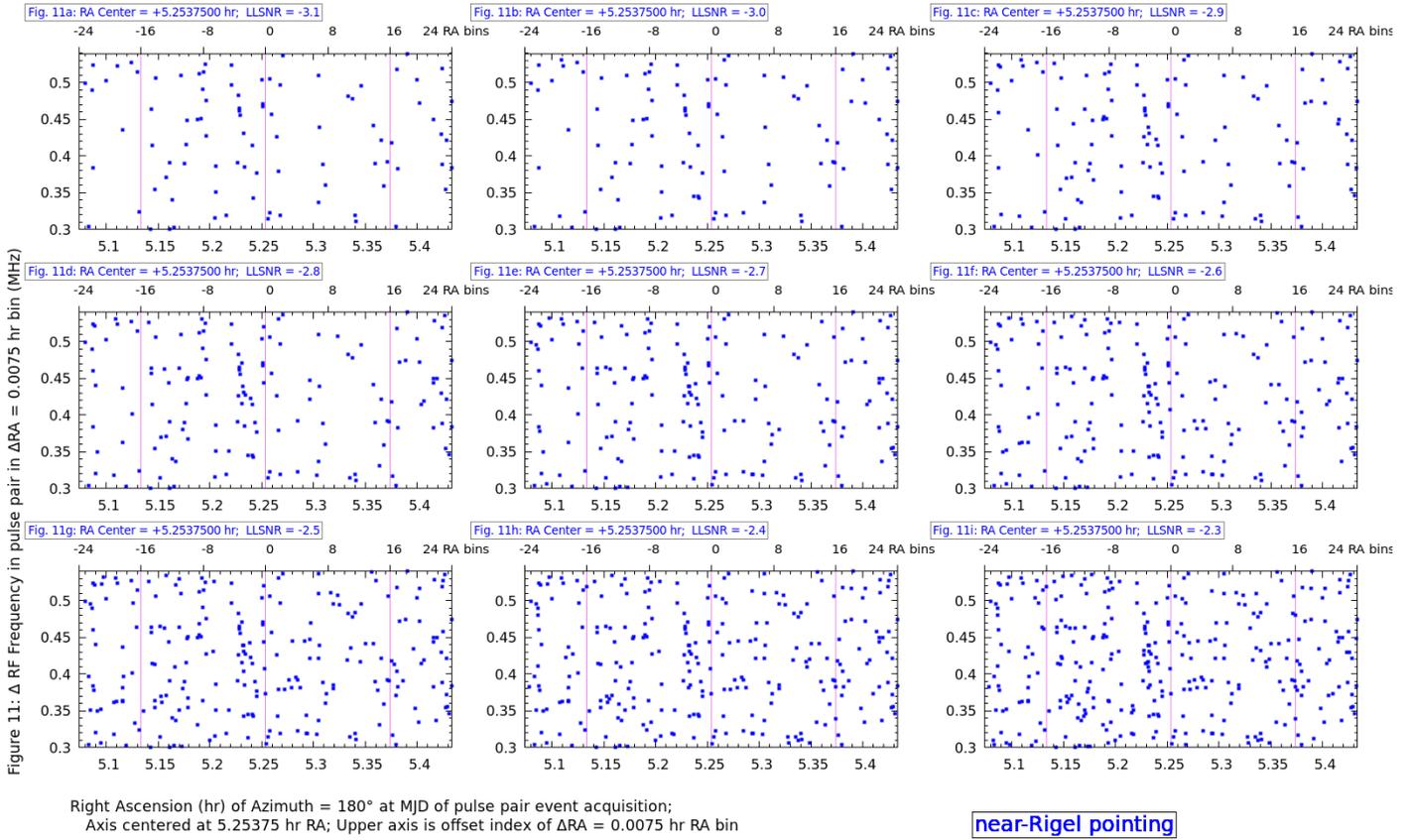

**Figure 11:** An anomalous concentration of pulse pair Δf measurements, evident in the direction of the near-Rigel DOI, was found after the DOI was identified, based on high pulse pair count statistical power and phase measurements indicating a celestial source. The Δf filter in second-level processing was then set to encompass these anomalies, and was used in all figures. The anomalies occur close to the RA region of high numbers of standard deviations of pulse pair count. A test using a 0.1 to 1 MHz Δf filter resulted in the reduction of the 6.2σ and 6.5σ peaks in **Fig.3** to peaks at 3.8σ and 4.9σ. Concentrations of pulse pair Δf values may be a method of identifying an intentional transmitting source, and provide a means of synchronizing symbols and decoding signals to information.

## V. DISCUSSION

Categories of measurement evidence in the 123 day duration experiment support an idea that the observed pulse pairs are celestially sourced, and do not appear to be explained by the expected emissions of natural celestial objects.

**Phase measurements support a celestial model**

**Fig. 2** indicates interferometer phase measurements that fit the expected azimuth of the prior near-Rigel DOI, due to the motion of the Earth during 123 days.

**Statistical power of pulse pair count**

In **Figs. 3-4,** statistical power of pulse pair count is observed near the center of the RA range reported in the prior experiments, examining the near-Rigel DOI. **[1]-[9]**

**Simultaneous wide bandwidth noise power increase is not observed**

**Figs. 5-8** show measurements of simultaneous power in two bandwidths, wider than the 3.7 Hz signal bandwidth, and do not indicate increasing power comparable to the 8.5 dB narrowband SNR candidate threshold, implying that natural objects are not the cause of the 3.7 Hz bandwidth anomalies.

Natural astronomical electromagnetic emission events are generally chaotic, having emission bandwidth much greater than 3.7 Hz. A high SNR 3.7 Hz bandwidth spectral component from a natural object emission can occur. If the broadband energy has characteristics of AWGN, SNR measurements will depend on Rayleigh amplitude statistics. Exceptions may exist after measurements are made. If the emitter has a power flux level well above the system noise, there could be a measurement of varying power spectral density over time and/or frequency. Non-flat noise, within a 954 Hz spectral segment, or within 50 MHz, and/or time-varying spectral power density, can cause the noise power measurement in the SNR calculation to deviate from the constant AWGN condition, producing narrow-bandwidth spectral outliers. Such variations may result, for example, due to the difference in the fluctuations of noise measurements between receiver system noise, and noise fluctuations that are wind-induced variable reflections and scattering from objects surrounding the interferometer antennas, and/or from mechanical movement of the antennas.

An example of natural object SNR-outlier pulse pairs was observed when 90 Jy natural object NRAO 5690,

contributed a 0.4 dB increase above system noise in a 50 MHz bandwidth. **Fig. 21** in **[9]** This finding implies that observation of natural objects having peak power density at 90 Jy or more, might produce a similar burst of continuum increase of 0.4 dB, and an increased number of pulse pairs. Such simultaneous increases of continuum in the near-Rigel DOI are not observed in **Figs. 7-8**. The development of natural object models having low Doppler spread and that might generate pulse pairs is ongoing work.

**MJD and RF Frequency measurements**

Measurement of RF Frequency is spread over wide ranges. A maser-cause would generally tend to exhibit consistent Doppler-shifted and Doppler-spread spectral lines. The presence of candidates at various MJDs imply a consistent emission, in contrast to a rare energy burst. **Figs. 9-10**

**Pulse pair frequency spacing**

Pulse pair frequency separation, concentrated in the range of 300 to 540 kHz, is unexpected given a wide-band AWGN-like source emission hypothesis. **Fig. 11.**

The presence of concentrated $\Delta f$ pulse pairs is expected in a high capacity communication system that is designed to be readily discovered and decoded. Further work is required to study natural object explanations.

## VI. CONCLUSIONS

The hypothesis predicting the possible existence of interstellar communication using pulse pairs has not been falsified in this work. Experimental results support a hypothesis of a celestial cause of the near-Rigel pulse pair anomalies. A natural object explanation of anomalies is conditionally falsified based on the absence of high power spectral density simultaneous measurements. Many alternate and auxiliary hypotheses are possible.

## VII. FURTHER WORK

1. Continue the observation and study of natural objects that might explain the pulse pair measurement anomalies.
2. A third interferometer element is under construction.
3. Test and improve the interferometer system. Seek equipment issues.
4. Perform another long duration interferometer experiment.
5. Seek independent observation of pulse pair anomalies.

## VIII. ACKNOWLEDGEMENTS